\documentclass{PoS}

\usepackage[utf8]{inputenc}
\usepackage[T1]{fontenc}
\usepackage{amsmath}
\usepackage{amssymb}
\usepackage{graphicx}
\usepackage{slashed}
\usepackage[small,bf]{caption}
\usepackage{upgreek}
\usepackage{cite}
\usepackage{dsfont}

\usepackage{epstopdf}

\newcommand{\be}{\begin{equation}}
\newcommand{\ee}{\end{equation}}
\newcommand{\ba}{\begin{eqnarray}}
\newcommand{\ea}{\end{eqnarray}}
\newcommand{\eq}{Eq.~}

\newcommand{\fig}{Fig.~}

\newcommand{\Lcal}{\ensuremath{\mathcal{L}}}
\newcommand{\Tr}{{\rm Tr}\,}
\newcommand{\lk}{\left[}
\newcommand{\rk}{\right]}

\def\lsi{\raise0.3ex\hbox{$<$\kern-0.75em\raise-1.1ex\hbox{$\sim$}}}
\def\gsi{\raise0.3ex\hbox{$>$\kern-0.75em\raise-1.1ex\hbox{$\sim$}}}
\newcommand{\lsim}{\mathop{\lsi}}

\title{
{\vspace{-25mm} \normalsize\hfill{\small MS-TP-09-28}\\\hfill{\small BI-TP-2009/29}}\\[20mm]
Screened perturbation theory for 3d Yang-Mills theory and the magnetic modes of hot QCD}

\ShortTitle{Screened perturbation theory}

\author{\speaker{Owe Philipsen}\\
        Institut f\"ur Theoretische Physik, 
        Westf\"alische Wilhelms-Universit\"at M\"unster, 48149 M\"unster, Germany \\
        E-mail: \email{ophil@uni-muenster.de}}

\author{Daniel Bieletzki\\
        Institut f\"ur Theoretische Physik, 
        Westf\"alische Wilhelms-Universit\"at M\"unster, D-48149 M\"unster, Germany \\
        E-mail: \email{Daniel.Bieletzki@uni-muenster.de}}

\author{York Schr\"oder\\
             Fakult\"at f\"ur Physik, Universit\"at Bielefeld, D-33501 Bielefeld, Germany\\
             E-mail:\email{yorks@physik.uni-bielefeld.de}}
             
\abstract{Perturbation theory for non-abelian gauge theories at finite temperature
is plagued by infrared divergences which are caused by magnetic soft modes 
$\sim g^2T$, corresponding to gluon fields of a 3d Yang-Mills theory. 
While the divergences can be regulated by a dynamically 
generated magnetic mass on that scale, the gauge coupling
drops out of the effective expansion parameter requiring summation of
all loop orders for the calculation of observables.
Some gauge invariant possibilities to implement such
infrared-safe resummations are reviewed. We use a scheme based on the 
non-linear sigma model to estimate some of the contributions $\sim g^6$ of the soft 
magnetic modes to the QCD pressure through two loops. The NLO contribution
amounts to $\sim 10\%$ of the LO, suggestive of a reasonable convergence
of the series.
}

\FullConference{International Workshop on QCD Green's Functions, Confinement, and Phenomenology - QCD-TNT09\\
                 September 07 - 11 2009\\
                 ECT Trento, Italy}

\begin{document}

\section{Introduction}

\begin{figure}[t]
\vspace*{-3cm}
\begin{center}
\includegraphics[width=\textwidth]{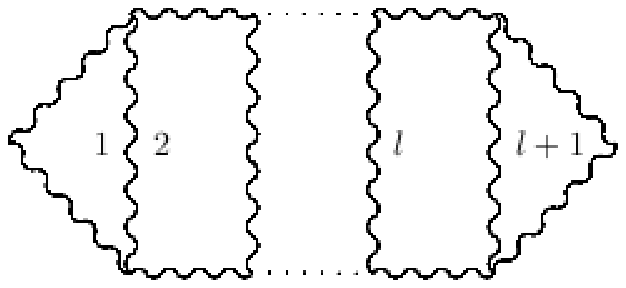}
\end{center}
\vspace*{-14cm}
\caption[]{Feynman diagram contributing to the pressure.}
\label{diagLinde}
\end{figure}
Perturbation theory for static quantities of non-abelian gauge theories at finite temperature (T) features three
scales: $\pi T$ associated with the non-zero Matsubara modes that arise due to the compactness 
of the Euclidean time direction, and $gT, g^2T$, where $g$ is the gauge coupling, which are associated with the screening 
of colour-electric and magnetic fields, respectively. A prohibitive obstacle for the evaluation of 
perturbative series are the well-known infrared divergences due to the magnetic modes.
For example, the thermodynamic pressure corresponds to all vacuum loop diagrams evaluated with 
finite temperature Feynman rules, and one $(l+1)$-loop example is shown in \fig\ref{diagLinde}. 
The Matsubara sum also contains a bosonic zero mode, and its contribution to the pressure is, parametrically,   
\be
\sim g^{2l}\left(T\int d^{3}p\right)^{l+1}p^{2l}(p^{2}+{m}^{2})^{-3l} \sim g^{6}T^{4}(g^{2}T/{m})^{l-3} \mbox{for} \quad l>3,
\ee 
where we have introduced a mass as infrared regulator in the propagators. Clearly, for 
$m\rightarrow 0$ the contribution diverges. One may argue that the full theory dynamically 
generates a magnetic mass scale, $m=const. g^2T$, to regulate this divergence. After all, we know that
non-perturbatively the theory is finite. However, in that case we see that the coupling constant
drops out of the expansion parameter and {\it all} higher loop orders contribute to the order $g^6$
in the pressure. This is the Linde-problem \cite{linde}, which occurs sooner or later
in any perturbatively computed observable, e.g.~in the Debye mass already at NLO, no matter how
weak the coupling. Thus, at
finite temperature, perturbation theory is only well-defined to some observable-dependent low order.
Note that this IR-problem in the zero-mode sector is equivalent to the one in 3d Yang-Mills theory.
 
In this contribution we reconsider a resummation method for 3d 
Yang-Mills theory, viz.~the finite T zero mode sector, which has been 
developed a while ago \cite{bp,bp1}. In view of current and future heavy ion collision experiments there 
is an urgent need for non-perturbative theoretical predictions. On the other hand, straightforward Monte Carlo
simulations of lattice QCD do not work for finite baryon densities or dynamical problems involving
real time (for a recent review, see e.g.~\cite{oprev}), motivating also analytical attempts. 
As the simplest observable to test our 
resummation scheme, we therefore consider the contribution of the infrared sector represented by 
the 3d Yang-Mills theory to the thermodynamic pressure of the QCD plasma. 
Of course, the equation of state
{\it can} be computed without major problems on the lattice, at least for a temperature range
$T\lsim 5T_c$. The idea here is to try and develop resummation schemes, test them against the pressure
and in case of satisfactory results apply them to other quantities. In the thermal context, resummation schemes
have already been successfully employed in scalar theories through four-loop level \cite{ja}, but 
it is not straightforward to generalise these methods to gauge theories. Recently a scheme
based on Hard Thermal Loop terms was attempted for non-abelian gauge theory, 
though not including the magnetic modes \cite{ja2}.
We therefore briefly review the contribution of the 3d gauge theory to 4d thermodynamics, before 
presenting our resummation scheme and evaluating it through two-loop order.

\section{High T effective theory from dimensional reduction}
\label{sec:dimred}

At sufficiently high temperatures, the previously mentioned scales get separated hierarchically, 
$g^2T\ll gT \ll \pi T$.
This suggests to successively integrate out those scales, which has been done systematically at the two-loop level. Here we follow \cite{kaj03}, to which we refer for details and references. 

In the first step, the hard non-zero Matsubara modes
are integrated out, leading to so-called EQCD as an effective theory for modes $\sim gT$ and softer,
\begin{eqnarray}
p_{QCD}(T)&=&p_{E}(T)+\frac{T}{V}\ln\int DA_{k}^{a}DA_{0}^{a}\exp{(-\int d^{d}x\mathcal{L}_{E})}\nonumber\\
\mathcal{L}_{E}&=&\frac{1}{2}Tr{F_{kl}^{2}}+Tr\left[D_{k},A_{0}\right]^{2}+{m_{e}^{2}}TrA_{0}^{2}+{\lambda_{E}^{(1)}}(TrA_{0}^{2})^{2}+{\lambda_{E}^{(2)}}TrA_{0}^{4}+...\;,\nonumber
\end{eqnarray}
where to leading order the parameters of the effective theory are 
$p_{E}\sim T^4;\quad m_{E}^{2}\sim g^2 T^2;\quad g_{E}^{2}\sim g^2 T;\quad\lambda_{E}^{(1)}\sim g^4 T;\quad\lambda_{E}^{(2)}\sim g^4 T$. Note that by performing the Matsubara sum the theory has
become effectively three-dimensional, and the electric gauge field component now represents an adjoint scalar field. The dots represent higher dimension operators that are suppressed by inverse 
powers of $T$.
This effective theory still contains the two dynamically generated scales 
$gT$ and $g^2T$. In a second step, the $A_0$ field and all modes living on the scale $gT$ can be
integrated out as well. This step already requires some resummation, but can still be performed as
a series in the coupling,
\begin{eqnarray}
\frac{T}{V}\ln\int DA_{k}^{a}DA_{0}^{a}\exp{(-\int d^{d}x\mathcal{L}_{E})}&=&p_{M}(T)
+\frac{T}{V}\ln\int DA_{k}^{a}\exp{(-\int d^{d}x\mathcal{L}_{M})}\nonumber\\
\mathcal{L}_{M}&=&\frac{1}{2}Tr{F_{kl}^{2}}+...\;,
\label{eq:L_M}
\end{eqnarray}
with matching coefficients $p_{M}\sim m_{E}^{3}T$ and $g_{M}^{2}\sim g_{E}^{2}$. This is an effective
theory for the ultra-soft modes $\sim g^2 T$, its leading term being equivalent to 3d Yang-Mills theory.
 
At this stage we can evaluate the contributions to the pressure 
from the different momentum scales separately and with appropriate methods. The contributions 
 $p_E(T)+p_M(T)$ can be computed as a series in the coupling constant, whereas the contribution
 of $p_M(T)$ is completely non-perturbative. This is the part of the theory giving the $g^6$ contribution
 to the pressure where the Linde-problem surfaces, and for dimensional reasons one can give its dependence
 on the coupling constant,
 \begin{eqnarray}
p_{G}(T)&=&\frac{T}{V}\ln\int DA_{k}^{a}\exp{(-S_{M})}\sim Tg^6\nonumber\\
\frac{p_{G}(T)}{T\mu^{-2\epsilon}}&=&d_{A}C_{A}^{3}\frac{g_{M}^{6}}{(4\pi)^{4}}\left[\alpha_{G}\left(\frac{1}{\epsilon}+8\ln\frac{\bar{\mu}}{2m_{G}}\right)+\beta_{G}+O(\epsilon)\right]\;.
\label{pressm}
\end{eqnarray}
The coefficients $\alpha_G,\beta_G$ first receive contributions at the four-loop level. 
There are similar (and more lengthy) expressions for $p_E(T)+p_M(T)$, for the full
result see \cite{kaj03}. 
It is well-known that whatever renormalisation is necessary and sufficient in the vacuum
will also be sufficient at finite temperature. The epsilon poles appearing in the contributions of 
the different scales are due to the break up of the momentum integration range, thus the sum of
all divergences has to cancel in the full pressure. 

Here our interest is merely in the contribution from the 3d Yang-Mills part.
Whereas $\alpha_G$ is known analytically \cite{kaj03}, the infrared coefficient 
$\beta_G$ is receiving contributions 
from all loop levels and needs to be evaluated non-perturbatively, e.g.~by lattice simulations of the 
effective theory \cite{betaG}, leading to a numerical value (with a certain error bar). 
In the following we discuss possibilities for an analytical evaluation by resummation
methods.

\section{Resummation schemes: general idea}

The general idea of a resummation is, at some given order in a perturbative scheme, to sum up higher order contributions (infinitely many in our case) into the current one. In order to avoid double counting,
these contributions must then be left out at the order where they naturally occur. In other words, the 
perturbative scheme gets reorganised in some systematic way. (For a discussion of various
schemes along those lines used in the context of thermal field theory, see \cite{bir}). 
This can be formalised by rewriting the Lagrangian serving for the perturbative expansion as \cite{jp} 
\be
\Lcal=\frac{1}{\ell}\left[\Lcal_{M}(\sqrt{\ell}X)+\Lcal_{\phi}(\sqrt{\ell}X)-\ell\Lcal_{\phi}(\sqrt{\ell}X)\right]\;,
\ee
where $X$ stands for a generic field;
$\Lcal_{M}$ is the theory under study, here taken as the 3d Yang-Mills theory of \eq(\ref{eq:L_M}); the modification
$\Lcal_{\phi}$ contains fields of the original Lagrangian and possibly auxiliary fields; and 
$\ell$ is a counting parameter introduced to systematise the resummed expansion, which is now an 
expansion in powers of $\ell$. At the end of a
calculation it is to be set to $\ell=1$, for which the Lagrangian is identical to the original one and the theory
is unchanged.
However, during a perturbative evaluation the contributions of $\Lcal_{\phi}$ are subtracted out
one order higher than they are added in, hence at any finite order of the expansion the result will
differ from the unresummed one. For example, the one-loop vacuum diagrams contributing to the pressure
will be of order $\ell^0$ while two-loop contributions count as $\ell^1$. In particular, if $\Lcal_{\phi}$
is chosen to represent a mass term for the gluon, this will regulate the infrared divergences.

Simply adding and subtracting a mass term for the gluon will not be sufficient, however, as we would obtain
gauge-dependent and hence arbitrary results at every finite order in our resummed scheme.
The question then is how to resum in a gauge-invariant way. The answer is that we need to resum
such that we maintain the Slavnov-Taylor identities of our gauge theory. Clearly, this is not just
to make things pretty, but a necessity in order not to change the original theory in the process of resumming.

\section{The non-linear sigma model providing a gluon mass term} 
 
A well-known and time honoured way to introduce mass for gauge fields is by means 
of the Higgs effect, i.e.~gauged sigma models \cite{corn}. 
For example, in the $SU(2)$-Higgs model used in the electroweak theory, we have
\begin{eqnarray}
\Lcal_{\sigma}&=&\Tr(D_{i}\Phi)^{\dag}(D_{i}\Phi)+\mu^{2}\Tr(\Phi^{\dag}\Phi)+2\lambda \Tr(\Phi^{\dag}\Phi)^{2}\\
\phi&=&\frac12(\sigma+i\tau^{a}\pi^{a})\quad\mbox{with}\quad T^{a}=\frac{\tau^{a}}{2}\nonumber\\
\sigma^{2}&=&v^2+(\pi^{a})^2\quad\mbox{with}\quad<\phi>=v\nonumber \;.
\end{eqnarray}
The gauge field gets a mass $m=g_M v/2$ through the expectation value of the scalar field.

However, in the current context we don't want an additional physical scalar particle, so
we decouple the $\sigma$ by sending
$\mu,\lambda\rightarrow\infty$ while keeping $v$ constant, which takes us to the non-linear sigma
model. Choosing a parametrisation by unitary matrices generalises to $SU(N)$ and we have
\be
\Lcal_{\phi}=\Tr [(D_i\phi)^\dag D_i\phi],\quad \phi(x)=\frac{m}{g_M}\exp(i\frac{g_M}{m}\pi^{\prime a}T^a),\quad \pi^{\prime a}\rightarrow\pi^a+O(\pi^3,\pi^5,...),\;
\ee 
with the usual covariant derivative $D_i=\partial_i-ig_MA_i$.
What is the meaning of the remaining scalar degrees of freedom? This can be understood
by considering the resummed partition function \cite{jp}
\be
Z=\int DA D\phi\;\Delta_{FP}\exp\lk-\frac{1}{\ell}(S_{YM}+S_\phi-\ell S_\phi + S_{gf})\rk\;,
\ee
with some gauge fixing that only depends on the gauge field, and the corresponding Faddeev-Popov
determinant $\Delta_{FP}$. We can now transform to unitary gauge by $A_i\rightarrow A_i^U$
with $U=\exp(i\pi^aT^a)$ and integrate over the auxiliary field. Up to a delta-function this gives
\be
\int D\phi \;\Delta_{FP}\,e^{-S_{gf}}=1\;,
\ee
such that we end up with 
\be
Z=\int DA\;\exp\lk-\frac{1}{\ell}(S_{YM}-m^2\int \Tr A^2 + \ell m^2\int \Tr A^2)\rk\;,
\ee
which is just pure gauge theory with a gauge invariantly resummed mass term. 
Unitary gauge is not well suited for perturbative higher order calculations
because of the bad UV-behaviour of gauge field propagators. Hence we shall
proceed in a slightly different way later on, but
these manipulations illustrate that the auxiliary field is merely a gauge degree of freedom and does
not add anything undesired to the theory.

Obviously, the choice of the resummation term $\Lcal_{\phi}$ above is not unique.
There are other choices compatible with gauge invariance that have been considered in the past,
like $\Lcal_{\phi}=F_\mu\frac{1}{D^2}F_\mu$, with $F_\mu=\epsilon_{\mu\alpha\beta}F_{\alpha\beta}$ 
\cite{jp}. This term does not require an auxiliary field, but at the price of being non-local.
Another non-local choice is based on the Chern-Simons eikonal and appears naturally in the 
framework of Hard Thermal Loop resummations \cite{an}. Yet another possibility to give mass
to the gauge field in a gauge invariant way is by the pinch technique \cite{cor97}.
Different gauge invariant additions/subtractions correspond to different ways of resumming the theory,
and  it is difficult to say a priori which ones are better or worse. Moreover, there is no small expansion 
parameter here. Writing $m=C g_M^2=C g^2T$, 
the dimensionless expansion parameter (modulo factors of $1/(4\pi)$) is 
\be
\frac{g_M^2}{m}=\frac{g_M^2}{Cg_M^2},
\ee
where $C$ is a numerical coefficient that will be discussed in the next section
(see also Table \ref{mmag}).
The coupling constant drops out and we are effectively expanding in a dynamically generated number.
Thus, there is no limit of parameters (not even $g_M\rightarrow 0$) in which the scheme is guaranteed to work. The convergence properties are due to the dynamics of the theory and can only be seen 
empirically by computing several orders of a given quantity.  

\section{Self-consistency and the magnetic mass}

\begin{table}[t]
\begin{center}
\begin{tabular}{|c|cc|cc|}
\hline
               & ref. && $m/g_M^2$& \\ \hline
1-loop gap eq. & \cite{an} &   & 0.38 & \\
               & \cite{bp,jp}& & 0.28 &\\
               & \cite{cor97}  & & 0.25 &\\
2-loop gap eq. & \cite{eb}   & & 0.34 &\\ \hline
lattice Landau & \cite{kar} & & 0.456(6)&\\ \hline
\end{tabular}
\end{center}
\caption[]{\label{mmag}
 Comparison of magnetic mass values for the pole mass of transverse gluons in SU(2), $m=C g_M^2$,
as obtained from gap equations and from gauge fixed lattice simulations.}
\end{table}

A priori the coefficient $C$ for the infrared cut-off is not determined. Since it gets subtracted
out again at higher orders in the resummation scheme it could, e.g., be treated as a variational parameter
with some freedom to optimise the expansion.
On the other hand, the resummation scheme of choice can be used to compute any observable, 
in particular one might
consider the self-energy of the gluon. Defining the ``magnetic mass'' to be the pole mass of the 
transverse part of the full gluon propagator, we obtain a gap equation for $m$, 
\begin{eqnarray}
D_{\rm trans}(p^2)=\frac{1}{p^2+m^2-\Pi_{\rm trans}(p^2)}\sim\frac{1}{p^2+m^2}\quad\mbox{for}\quad p^2=-m^2\nonumber\\
\Pi_{\rm trans}(p^2=-m^2)\left(1+\frac{\partial\Pi_{\rm trans}}{\partial p^2}(p^2=-m^2)\right)=0\nonumber
\end{eqnarray}
Note that the pole of the transverse self energy is gauge-invariant order by order, i.e.~this definition
of $m$ is gauge invariant. Results for the solution of the gap equation using 
various resummation 
schemes and gauge fixed lattice propagators 
are summarised in Table \ref{mmag}. Expectedly there is some scatter in the values,
but it is not much more than the two-loop correction computed with the scheme based on 
the non-linear sigma model \cite{eb}, which amounts to about 15\%.
This suggests that the schemes might possibly lead to a reasonable convergence.

\begin{figure}[t]
\begin{center}
\vspace*{0.3cm}
\includegraphics[width=0.5\textwidth]{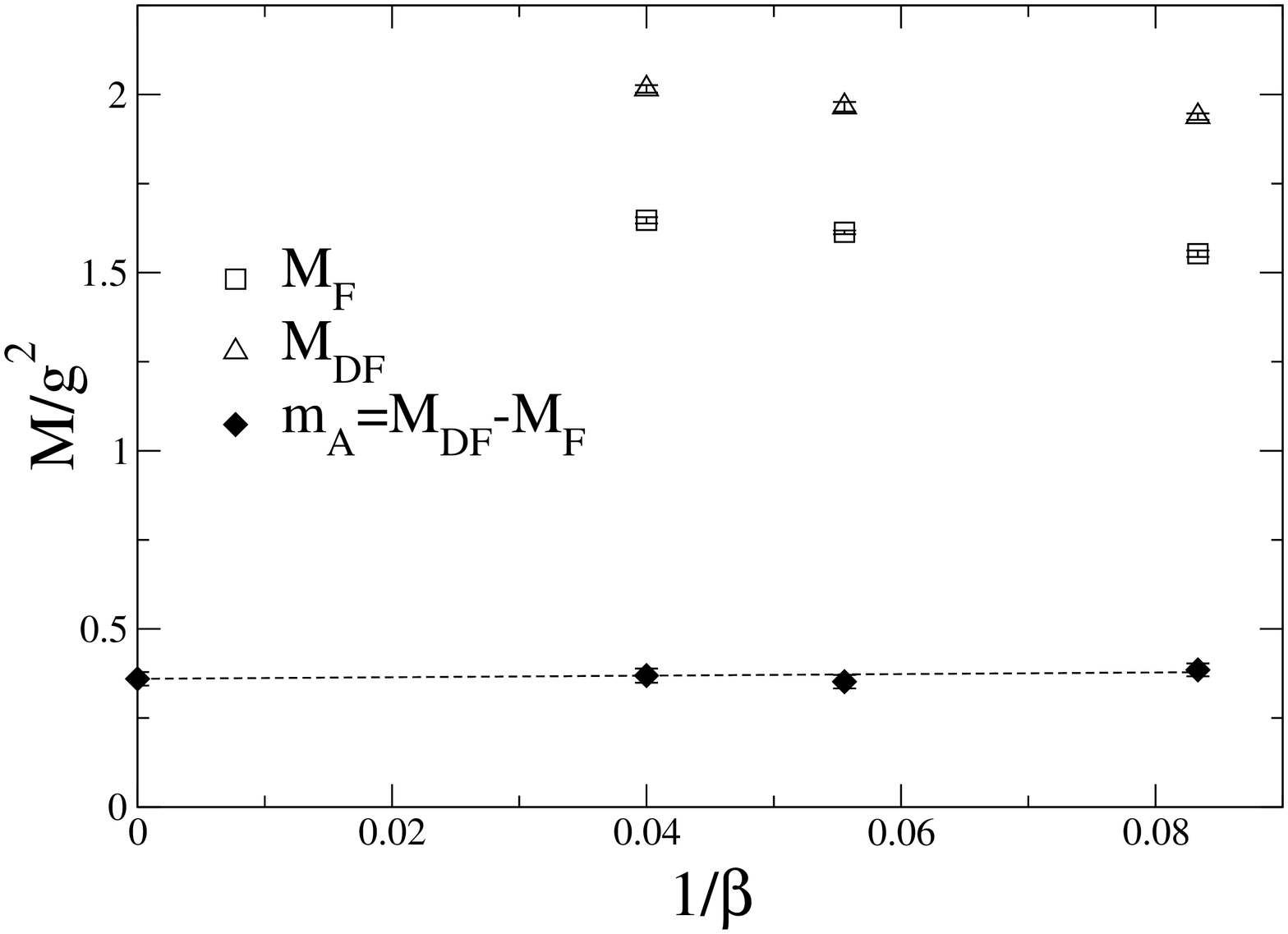}
\end{center}
\vspace*{-0.3cm}
\caption[]{Lowest masses from the exponential fall-off of the field strength correlators in 
\eq(\ref{operator}). Both masses are diverging at the same rate, but the difference can be extrapolated to
the continuum.}
\label{mm}
\end{figure}

There is also an evaluation of the magnetic mass in a gauge invariant 
lattice calculation, based on the following observable \cite{op}
\be
\frac{\langle (D_iF_{ij})^a(x)U^{Ad}_{ab}(x,y)(D_kF_{lm})^b(y)\rangle}{\langle (F_{ij})^a(x)U^{Ad}_{ab}(x,y) (F_{lm})^b(y)\rangle}\sim \exp[-(3m-2m)|x-y|]\;,
\label{operator}
\ee
where $U^{Ad}_{ab}(x,y)$ is a Wilson line in the adjoint representation. In perturbation theory,
the correlator in the numerator is at large distances dominated by three-gluon exchange, whereas
the denominator is dominated by two-gluon exchange, hence the ratio should fall off with one 
gluonic pole mass. In a non-perturbative reasoning these correspond to gluelump correlators,
i.e.~glue with certain quantum numbers bound to a static adjoint source. The ratio of these  correlators
measures the smallest excitation energy with the quantum numbers of a gluon. Note that, due to the diverging self energy of the adjoint Wilson lines, neither gluelump correlator has a continuum limit, but the ratio does and the energy difference is finite.
The resulting continuum extrapolation is shown in \fig\ref{mm}, and gives $m=0.36(2)g_M^2$.

Finally, the mass parameter in the gluon propagator is not a physical observable and not the goal of our
investigation, but merely a quantity that appears as a regulator for our resummation scheme.
The test of the latter is in computing a physical observable. In the context of thermal field theory, this has
been done by using the linear sigma model as a dimensionally reduced version of the electroweak theory
in order to calculate the electroweak phase transition as 
a function of the Higgs mass. Standard perturbation theory to leading order predicts a
first order phase transition for all Higgs masses, and cannot be extended beyond LO because of the
massless W-bosons in the symmetric phase. Using the resummation similar to the one described here,
the critical Higgs mass for which the electroweak phase transition disappears and turns into a smooth
crossover was predicted \cite{bp, bp2} and is within 10\% of the corresponding lattice 
results \cite{kaj, fodor}. 

\section{Application to the pressure}

\begin{figure}[h!]
\begin{center}
\includegraphics[width=1.05\textwidth]{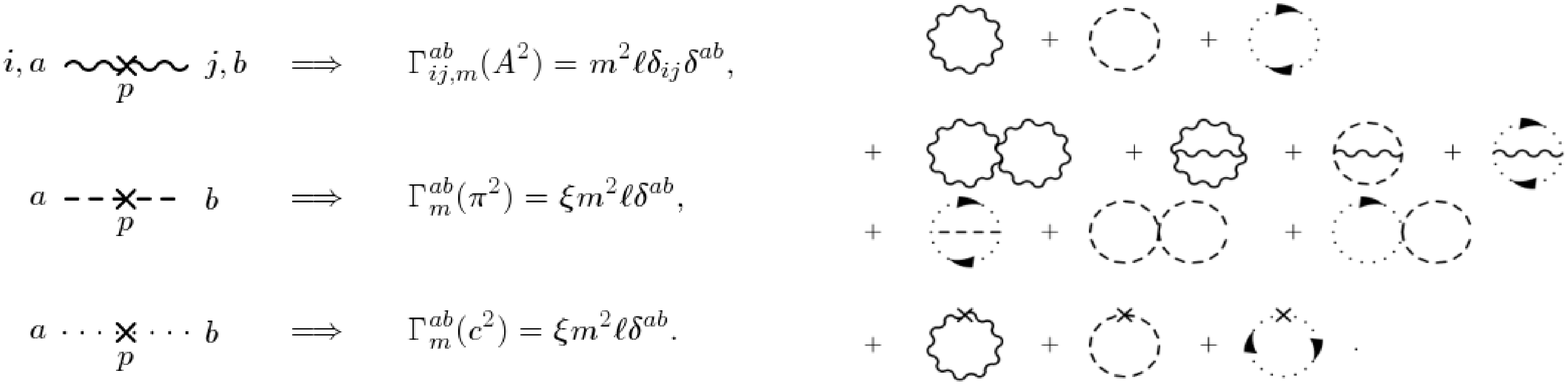}
\end{center}
\caption[]{Left: New counterterms induced by the effective Lagrangian
  \eq(\ref{eq:Leff}). Right: Feynman diagrams for the contribution of magnetic QCD to the pressure at
  order $\ell$.}
\label{diag}
\end{figure}

Let us now apply the screened perturbation theory based on the non-linear sigma-model to the 
calculation of the pressure of the soft magnetic field modes, \eq(\ref{pressm}).
In the framework of the resummation outlined above, the Lagrangian in the exponent of the first line
is supplemented by the sigma-model term, the counter term as well as gauge fixing and ghost terms.
How many powers of the auxiliary fields $\pi^a(x)$ are needed depends on the order of the calculation.
At the two loop level, or to order $\ell^1$, a four-point vertex is the highest that is needed, and we have
\begin{alignat}{5}
\Lcal_{\rm eff}=&\,\frac14\left(\partial_{i}A_{j}^{a}-\partial_{j}A_{i}^{a}\right)^2+\frac{1}{2\xi}\left(\partial_{i}A_{i}^{a}\right)^2+\frac12 m^2 A_{i}^{a}A_{i}^{a}\nonumber\\
&\,+\frac12\left(\partial_{i}\pi^{a}\right)^2+\frac12\xi m^2\pi^{a}\pi^{a}+(\partial_{i}(c^{a})^{*})\partial_{i}c^{a}+\xi m^2 c^{a\text{*}}c^{a}\nonumber\\
&\,+g_M\sqrt{\ell}f^{abc}A_{i}^{b}A_{j}^{c}\partial_{i}A_{j}^{a}+\frac14 g_M^2 \ell
f^{abe}f^{cde}A_{i}^{a}A_{j}^{b}A_{i}^{c}A_{j}^{d}\nonumber\\
&\,+\frac12 g_M\sqrt{\ell}f^{abc}\left(\partial_{i}\pi^{a}\right)A_{i}^{b}\pi^{c}+g_M\sqrt{\ell}f^{abc}(\partial_{i}(c^{a})^{*})A_{i}^{b}c^{c}-\frac12 g_M\sqrt{\ell}\xi mf^{abc}\pi^{a}c^{b}(c^{c})^{*}\nonumber\\
&\,+\frac{1}{8}\frac{g_M^2 \ell}{m^2}\left(\frac{2}{N}\delta^{ab}\delta^{cd}+d^{abe}d^{cde}\right)\pi^{a}\pi^{c}(\partial_{i}\pi^{b})\partial_{i}\pi^{d}\nonumber\\
&\,-\frac18 g_M^2 \ell\xi\left(\frac{2}{N}\delta^{ab}\delta^{cd}+d^{abe}d^{cde}-f^{abe}f^{cde}\right)(c^{a})^{*}\pi^b\pi^c c^{d}\nonumber\\
&\,-\frac12 m^2 \ell A_{i}^{a}A_{i}^{a}-\frac12\xi m^2 \ell\pi^{a}\pi^{a}-\xi
m^2 \ell c^{a\text{*}}c^{a}.
\label{eq:Leff}
\end{alignat}
Note the additional Feynman rules that result from this effective Lagrangian. Besides the vertices
for the auxiliary fields and their interactions with gluons, ghosts and themselves, there are also
two-point vertices associated with the counter terms, cf.~\fig\ref{diag} (left), 
since they are treated as interactions.
Through order $\ell^1$, i.e.~two loops, we then have to compute the diagrams in \fig\ref{diag} (right).

Our calculation is performed in a general $R_\xi$-gauge. This is an excellent way of checking the
calculation -- the pressure as a physical quantity has to be independent of the gauge parameter
$\xi$. For intermediate checks, it is also useful to observe that the set of diagrams has three obviously
gauge invariant subsets. The one-loop diagrams from the first line, and the two-loop diagrams
just correspond to an unresummed non-linear sigma model which must be gauge invariant order by order.
Correspondingly, the counter term diagrams  in the last line are also a gauge invariant subset.
According to the reshuffling of the expansion, they appear only at order $\ell$ although they technically are one loop diagrams.

The results for these three sets of diagrams for $SU(N)$ are 
(using $g_M^2=g^2T$, cf.\ Section \ref{sec:dimred})
\ba
p_{G,\ell^0}(T)&=&C^3(N^2-1)\frac{g^6}{6\pi}T^4\nonumber\\
p_{G,ct}(T)&=&-\ell C^3(N^2-1)\frac{g^6}{4\pi}T^4\nonumber\\
\frac{p_{G,\ell^1}}{\mu^{-2\epsilon}T^4}&=&\ell C^2N(N^2-1)\frac{g^6}{(4\pi)^2}
\left[-\frac{21}{64}\left(\frac{1}{\epsilon}+4\ln\frac{\bar{\mu}}{2Cg^2T}\right)
+\frac{3}{16}-\frac{21}{32}-\frac{21}{16}\ln\frac{2}{3}+O(\epsilon)\right]\;.
\ea
Because of the resummation, we are now getting a contribution of order $g^6$ already
to leading order in this scheme, whereas in unresummed perturbation theory this would
occur at four loops. We then extract the following coefficients 
(with numerical values for $\ell=1$, $N=2$),
\ba
\alpha_G&=&-\frac{21}{4}\frac{C^2}{N^2}\pi^2\ell=-1.015582 \nonumber\\
\beta_{G,\ell^0}&=&\frac{128}{3}\frac{C^3}{N^3}\pi^3=3.630132\nonumber\\
\beta_{G,ct}&=&-64\ell\frac{C^3}{N^3}\pi^3=-5.445198 \nonumber\\
\beta_{G,\ell^1}&=&-\left(\frac{15}{2}+21\ln\frac{2}{3}\right)\ell\frac{C^2}{N^2}\pi^2=0.196301
\label{betas}
\ea
In order to evaluate these contributions, a number for $C$ had to be specified.
To be fully self-consistent, we have taken this to be the magnetic mass evaluated in the 
same resummation scheme for the case of $SU(2)$, i.e.~$C\approx 0.28$ \cite{bp}.  
We now have two consecutive orders for $\beta$ and may get a first glimpse of the convergence 
properties. While the resummed perturbation is organised in orders of $\ell$, a convergence check
by comparing different orders in $\ell$ would not seem to make much sense. By construction, the counter
term diagrams get subtracted one order in $\ell$ higher than the other diagrams with the same number
of loops and the same integral structure. Hence, when asking for the convergence
properties, it would seem natural to proceed loopwise. Thus, in \eq(\ref{betas}) the contribution of all
one loop diagrams is $\beta_{G,\ell^0}+\beta_{G,ct}=-1.81$, while the genuine two-loop contribution is 
$\beta_{G,\ell^1}$, which is about  10\% correction. We would expect the counter terms with two loops,
entering at $l^2$-level,
to be of the same order of magnitude as the two-loop contributions evaluated here. Thus, comparing
the one-loop and two-loop results appears to be promising in terms of convergence.  
We are presently performing the three-loop computation in order to check this behaviour. 

Another interesting observation is the fact that if we take $C$ to be the coefficient 
of the magnetic mass, evaluated in the same resummation scheme, then $C\sim N$. Thus, $N$ drops
out of the expressions in \eq(\ref{betas}) and $\beta_G$ becomes $N$-independent, as expected on
general grounds \cite{kaj03}. 

\section{Conclusions}

We discussed the possibility to gauge-invariantly resum the 3d Yang-Mills theory such as to 
self-consistently include a dynamically generated pole mass for the gluons, which serves as a regulator
for infrared divergences. In such a scheme a perturbative expansion to arbitrary orders is possible.
However, the dimensionless expansion parameter is a dynamically generated number, hence there is no
parametric limit in which convergence of the series is guaranteed. Rather, the convergence properties
can only be inspected after the calculation of several orders. We have applied a resummation scheme
based on the non-linear sigma-model to evaluate the contribution of the soft magnetic gluons to the 
pressure of the QCD plasma. Comparison between one- and two-loop contributions suggests that
there is hope for a reasonable convergence of the series. 

\section*{Acknowledgements}
We thank H.~Malekzadeh for dicsussions and checks. 
This work is partially supported by the German BMBF, grant 06MS9150.

\end{document}